%% file: yjyang.tex
\documentstyle[jkas2]{article}
\runningauthor{Y. Yang, H. S. Park,  M. G. Lee, \&  S.-G. Lee}
\runningtitle{Near-IR Photometry of the Arches Cluster}

\beginpage{1}
\endpage{10}

\begin{document}

\title{Gemini Near-IR Photometry of the Arches Cluster near the Galactic Center}
\author{Yujin Yang,  Hong Soo Park,
Myung Gyoon Lee\footnote[1]{Corresponding Author: M. G. Lee}, and Sang-Gak Lee}

\address{Astronomy Program, SEES, Seoul National University, Seoul, 151-742,
 Korea \\
{\it E-mail: yjyang@astro.snu.ac.kr, hspark@astro.snu.ac.kr,
 mglee@astrog.snu.ac.kr and sanggak@astrosp.snu.ac.kr}}

\address{\normalsize{\it (Received ???. ??, 2002; Accepted ???. ??,2002)}}

\abstract{
\begin{center}\begin{minipage}{0.9\textwidth}
We present Near-IR photometry of the Arches cluster,
a young and massive stellar cluster near the Galactic center.
We have analyzed the high resolution (FWHM $\sim$ 0.2$''$)
$H$ and $K'$ band images
in the \emph{Galactic Center Demonstration Science Data Set},
which were obtained with the Gemini/Hokupa's adaptive optics (AO) system.
We present the color-magnitude diagram, the luminosity function and
the initial mass function (IMF) of the stars in the Arches cluster
in comparison with the $HST/NICMOS$ data.
The IMF slope for the range of $1.0< \log~ (M/M_\odot) <2.1$ is estimated
to be $\Gamma = -0.79\pm0.16 $, in good agreements with the earlier
result based on the $HST/NICMOS$ data [Figer et al. 1999, ApJ, 525, 750].
These results strengthen the evidence that the IMF of the bright stars
close to the Galactic center is much flatter than that for the solar
 neighborhood.
This is also consistent with a recent finding that the IMFs of the bright
 stars in
young clusters in M33
get flatter as the galactocentric distance decreases [Lee et al. 2001, astro-ph
 0109258].
It is found that the power of the Gemini/AO system is comparable,
with some limits,
to that of the $HST/NICMOS$. 
\end{minipage}\end{center}
}


\maketitle

\section{INTRODUCTION}

The Arches cluster is a very unique cluster in the Milk Way,
because it is a very massive and compact young cluster close
to the Galactic center.
It was confirmed as a star cluster  including emission-line stars by Nagata et
 al (1995).
To date only three clusters are known to be very close to the Galactic center.
The other two clusters are the Quintuplet cluster and the IRS 16 cluster at the
 Galactic center.
The size of the Arches cluster is about $15''$ (= 0.58 pc at the distance of 8 kpc),
and the total mass is estimated to be about $10^4 M_\odot$ (Figer et al 1999). 
The Arches cluster has a very high peak density $3 \times 10^5$ $M_\odot$
 pc$^{-3}$
in the inner $9''$ (0.35 pc), showing that it is one of the densest known young
 clusters
in the Local Group galaxies. Similar examples are R 136, the central cluster of
 30 Dor
in the Large Magellanic Cloud and NGC 3603 in our Galaxy.
The age of the cluster is estimated to be about 2--5 Myrs (Figer et al 1999,
Blum et al 2001).
Very recently  Yusef-Zadeh et al (2002) detected, using the Advanced CCD Imaging
Spectrometer on board Chandra X-Ray Observatory,  two X-ray sources in this
 cluster,
and suggested that the X-ray emission from the sources arises
from  stellar wind sources in the cluster.

The presence of compact young clusters
like the Arches cluster and the other two clusters near the Galactic
center indicates that stars are forming even in such a dense environment.
Therefore a study of these clusters will provide important hints for understanding
the star formation process under extreme environments.

Stars in the Arches cluster were studied in detail for the first time
by Figer et al (1999) and Kim et al (2000)
who used the  Hubble Space Telescope ($HST$) Near-Infrared Camera and
Multiobject Spectrometer ($NICMOS$) observations.
Figer et al (1999) found several interesting results on this cluster:
(1) the Arches cluster is very young, with an age of only about 2 Myrs, showing
that stars are forming very recently in the region close to the Galactic center;
and
(2) the initial mass function (IMF) of the massive stars
in this cluster is derived to be significantly flat, having a slope of
$\Gamma = \log N / \log M = -0.7\pm0.1$ for the mass range of $0.8 < \log ~(M/M_\odot) <2.1 $.
Surprisingly this IMF slope is much flatter than the average for other clusters in the solar
neighborhood which is close to the Salpeter value, $\Gamma= -1.35$
(see Scalo 1998).
This result shows, if confirmed, that stars with flatter IMFs are formed in the
dense region like the Galactic center, while stars with steeper IMFs are formed
in the low-density region like the solar neighborhood.

In spite of the importance of the study of these clusters, there are only a few studies
of these clusters to date.
It has been a demanding job to investigate the IMF
for the clusters like the Arches from the ground-based observation,
because the cluster fields are very crowded and the interstellar extinction toward
the clusters is severe. Therefore the $HST$ remains to be almost the only instrument useful
for these studies until recently.
However, with the advent of Adaptive Optics (AO) system at the Gemini Telescope,
it became possible to study the stars in the clusters like the Arches
in detail with ground-based observations.

      \begin{table}[!htb]
      \begin{center}
      {\bf Table 1.} Observation log \\
      \begin{tabular}{ c c c c c }
      \hline \hline
      ID & Date   & Filter  & Total exp.   & FWHM \\
         & (2000) &         &  time(sec)   & $('')$ \\
      \hline
      1 & 07-05 & $H$  & 3    & 0.140-0.165 \\
      2 & 07-05 & $H$  & 720  & 0.180-0.230 \\
      3 & 07-09 & $K'$ & 2    & 0.120-0.140 \\
      4 & 07-30 & $K'$ & 16   & 0.105-0.135 \\
      5 & 07-04 & $K'$ & 180  & 0.185-0.250 \\
      6 & 07-03 & $K'$ & 240  & 0.145-0.180 \\
      7 & 07-30 & $K'$ & 480  & 0.125-0.145 \\
      8 & 07-09 & $K'$ & 1020 & 0.125-0.135 \\
      \hline
      \end{tabular}
      \end{center}
      \label{obslog}
      \end{table}

In this paper we present Near-IR photometry of the Arches cluster
obtained for science demonstration using the Gemini/AO system
in comparison with the $HST/NICMOS$ results.
Preliminary results of this study were presented by Yang et al (2002).
During the preparation of this paper, Stolte et al (2002) also presented
at a conference a similar study to ours using the same data set.

\section{DATA SET and DATA REDUCTION}

\subsection{Data Set}

We have used the data set prepared as part of the
\emph{ Galactic center demonstration science}
which was obtained using Hokupa's Adaptive Optics system
at the Gemini/ North telescope in July, 2000.
We have analyzed the preprocessed data released to the public.
This data set contains deep, high resolution $H$ and $K'$ band (1.65,
2.12 $\mu m$) images of the central regions of the Arches cluster.
The pixel scale is $0''.02$/pixel, giving a total field of view
 $20''.5\times20''.5$.

Table 1 lists the observation log.
The FWHMs of the point sources are measured to be $\sim 0.2''$ in the $H$ band images,
and $\sim 0.1''$ in the $K'$ band images.
Fig. 1. displays gray scale maps of the Gemini images of the Arches clusters in comparison with the
$HST/NICMOS$ images.
The field of view of the Gemini images is smaller than that of the $NICMOS$
 images,
as shown in Fig. 1, but covering the most central region of the cluster.
Fig.1 shows that stars in the cluster are very well resolved in the Gemini images.

\subsection{Data Reduction}

We have derived the instrumental magnitudes of the stars in the images
using the point spread function fitting package of IRAF /DAOPHOT (Stetson 1987).
Because the median sky level was subtracted from the image in the released
data set, we added this value to the raw image to derive photometric errors. 

Because this data in the Gemini demonstration program was obtained
primarily  to check the performance
of the telescope and instruments and to test the data reduction
scheme, this data set has some limitations for detailed study of the Arches
cluster.
First, the exposure time of the $H$ images is not as deep as $K'$
images. Therefore fewer faint stars are detected in the $H$ images than in the $K'$ images.
Second, the field coverage of each exposure is somewhat different, reducing the
areas common among the images.
Third, the observation for standard stars is not available.
However, these images are still the sharpest and deepest images
ever taken of this cluster at the ground-based telescope.


Because the observation for standardization was
not available for this data set,
we have calibrated our photometry using the $HST/NICMOS$ data
given by Figer et al (1999).
We have transformed the
instrumental $H$ and $K'$ magnitudes into the $HST/NICMOS$ system
using the photometry given by Figer et al (1999).
The effective wavelengths of the $NICMOS$ filters F160W and F205W are,
respectively, 1.60 and 2.05 $\mu m$,  which are slightly different
from the Gemini filters.

The coordinate transformation equations between the Gemini photometry
and the $NICMOS$ photometry are derived as follows:
\[ X(NIC) = 0.187 X(Gem) -0.187 Y(Gem)+225.421, \] 
\[Y(NIC) = -0.189 X(Gem) -0.187 Y(Gem)+ 490.104 \]
where the coordinates are given in units of pixel of the $NICMOS$ and the
Gemini images, respectively.

Using the bright stars common between the Gemini and $NICMOS$ images
we have derived the transformation equations as follows:
\[ m_{\mathrm{F205W}}  = k + 0.103  ~(h-k) - 0.587, ~~\sigma = 0.058, \] 
\[ m_{\mathrm{F160W}}-m_{\mathrm{F205W}} = 0.924 ~(h-k) + 1.361, ~~\sigma = 0.062 \]
where $h$ and $k$ represent the instrumental magnitudes of Gemini.
Hereafter we use $H$ and $K$, respectively,
for the calibrated magnitudes in the Gemini photometry,
$m_{\mathrm{F205W}}$ and $m_{\mathrm{F160W}}$.
Fig. 2 shows the differences in magnitudes
and colors between  the $NICMOS$ and the Gemini for the stars used for the
transformation.
Relatively large scatters seen in Fig. 2 appear to be due to
the variation of the PSFs and the difference in the filters of the two systems.

Table A1 lists the photometry of the 327 measured bright stars
($K<17$) of the Arches cluster.
The $X$ and $Y$ coordinates listed in Table A1 are given in units of the pixel
which corresponds to 0.02 arcsec.  $X$ increases to the west and $Y$ increases
to the north in Fig. 1(left). Some bright stars are labeled for identification
in Fig. 1(left).

\section{Color-Magnitude Diagram}

We display a color-magnitude diagram (CMD)
of the measured stars in the Arches cluster in Fig. 3.
Fig. 3 shows that most of the stars in the Arches cluster
\begin{figure*}[tbhp]
\centerline{
  \epsfysize=0.45\textwidth\epsfbox{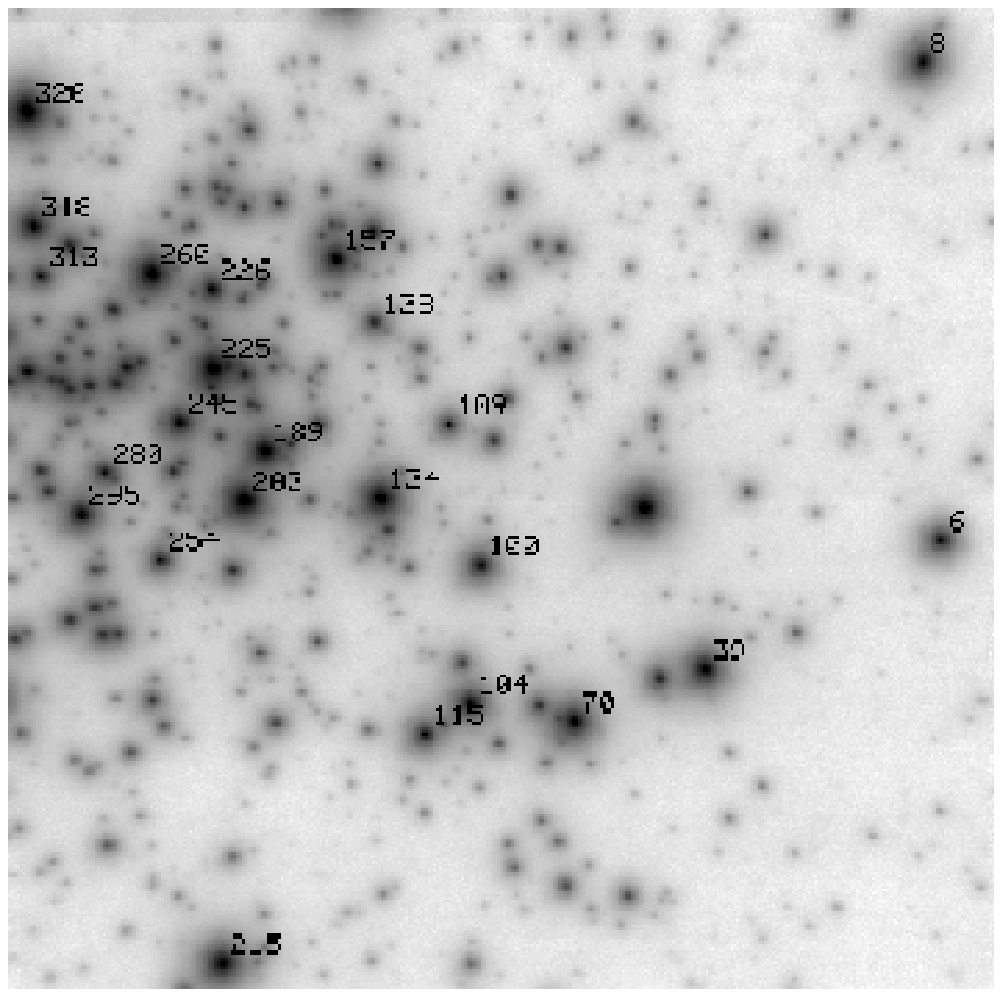}  
  \epsfysize=0.45\textwidth\epsfbox{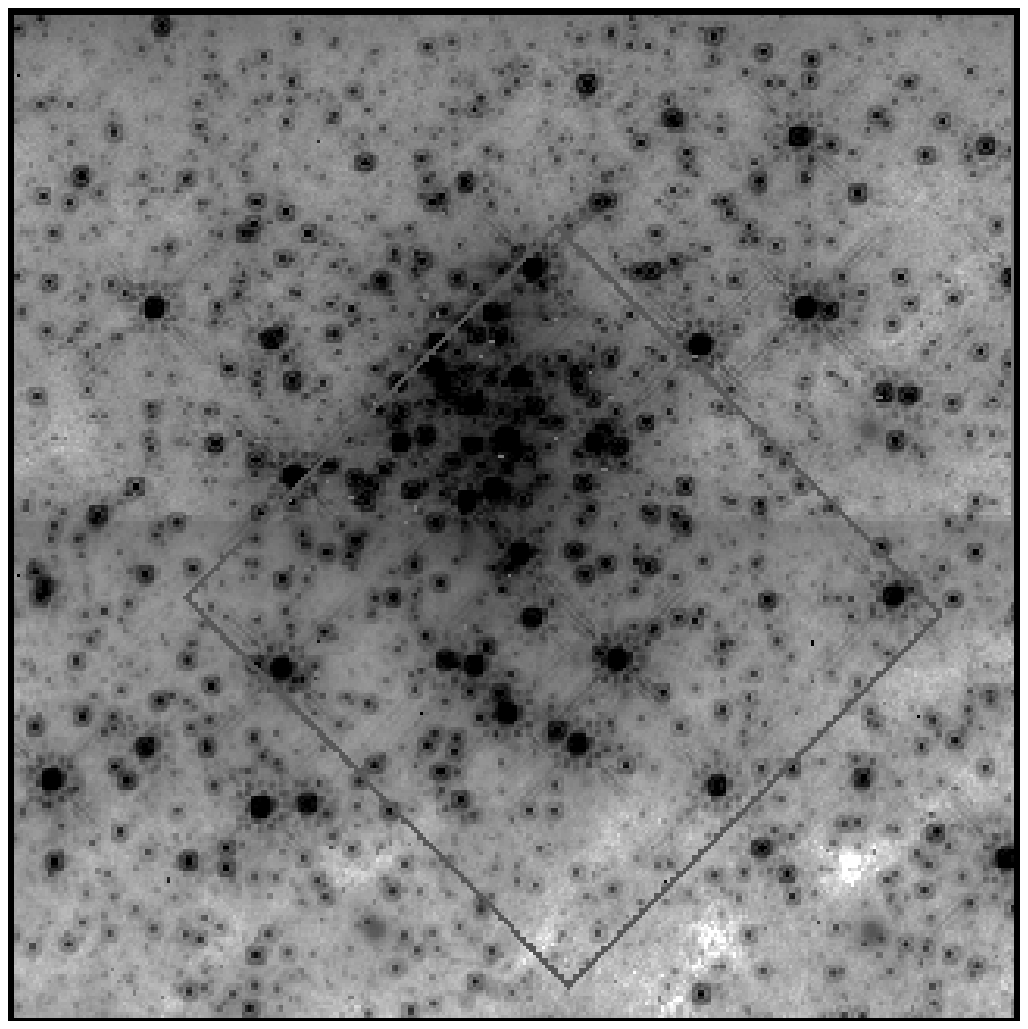}}   
\label{image}
\vskip 0.5cm
\begin{center}
\begin{minipage}{0.9\textwidth}
{\bf Fig. 1.}---~~(Left) A gray scale map of the Gemini $K'$ band image.
The size of the field of view is $20''.5\times20''.5$. 
North is up and east is to the left.
(Right) A gray scale map of the $HST/NICMOS$ $K(F205W)$ band image.
The solid line represents the Gemini field shown in the left.
\end{minipage}
\end{center}
\end{figure*}
\begin{figure*} [!bhtp]
\epsfxsize=0.90\textwidth\epsfbox{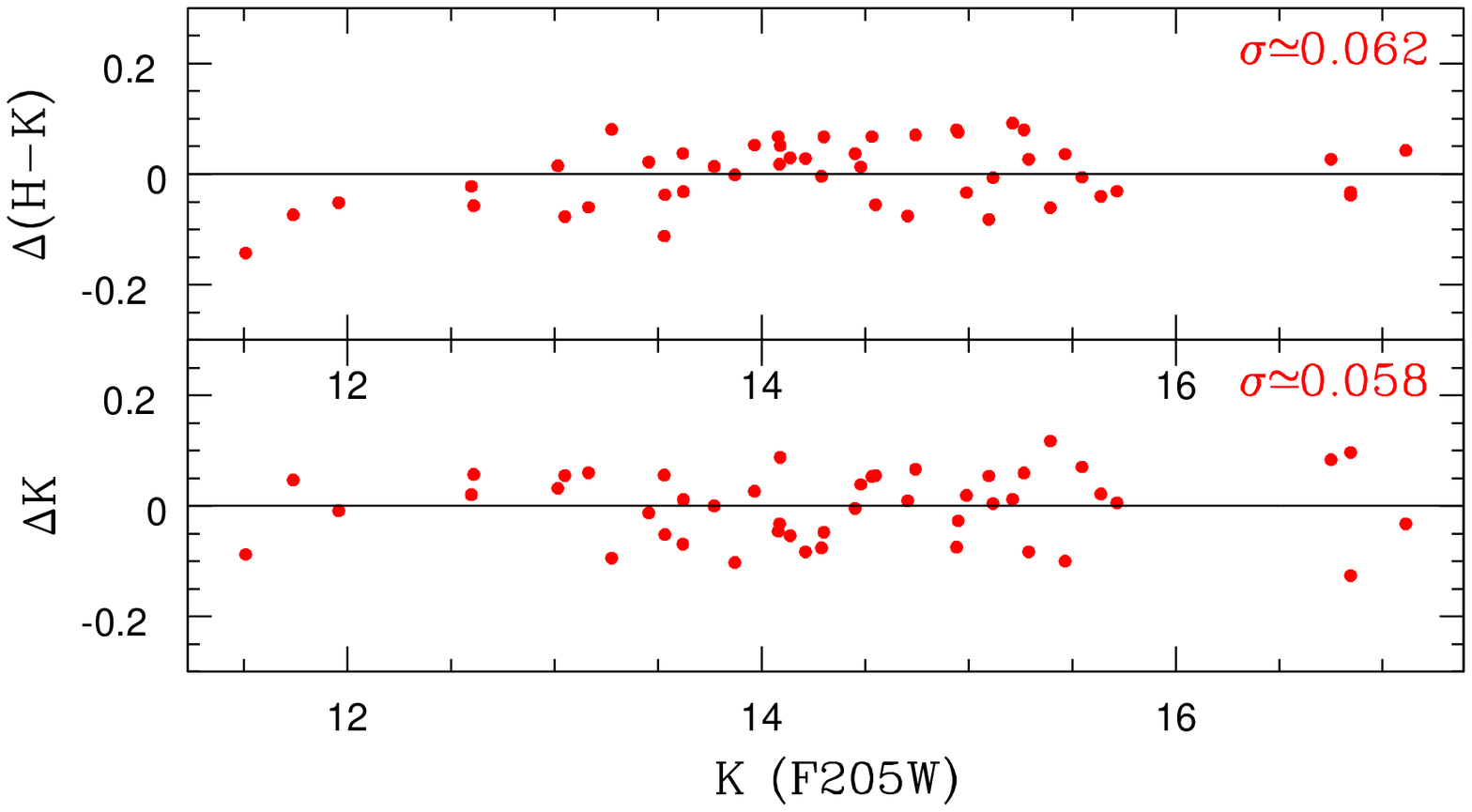} 
\vskip 0.5cm
\hskip 0.05\textwidth
\begin{minipage}{0.9\textwidth}
{\bf Fig. 2.}---~~The differences in magnitudes and colors between
the $NICMOS$ photometry and the Gemini photometry for bright stars.
The differences are given in terms of
$\Delta$ = $NICMOS$ photometry minus Gemini photometry.
\end{minipage}
\end{figure*}
\begin{figure*} [!htbp]
\centerline{
  \epsfysize=0.43\textwidth\epsfbox{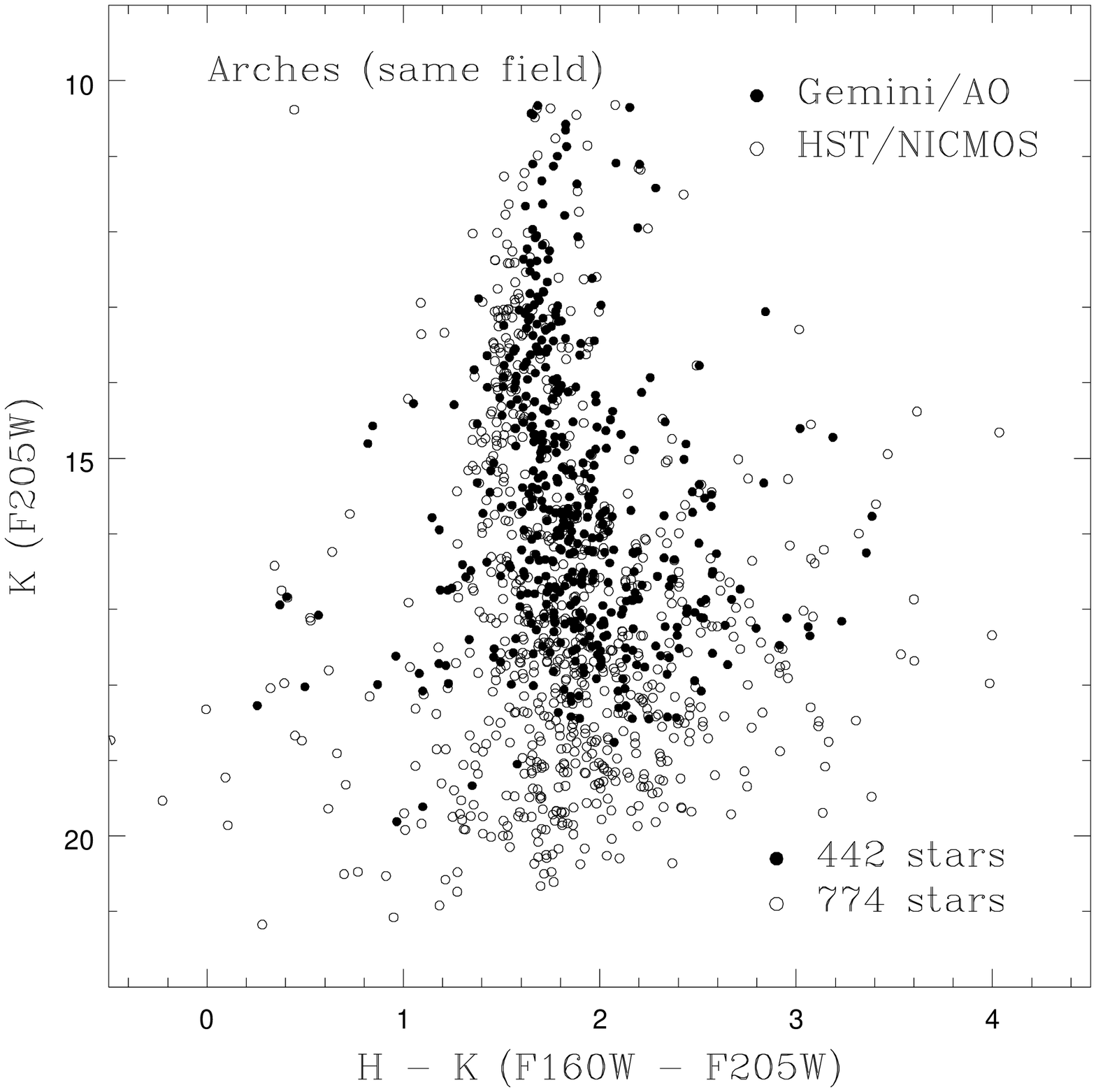} 
  \epsfysize=0.43\textwidth\epsfbox{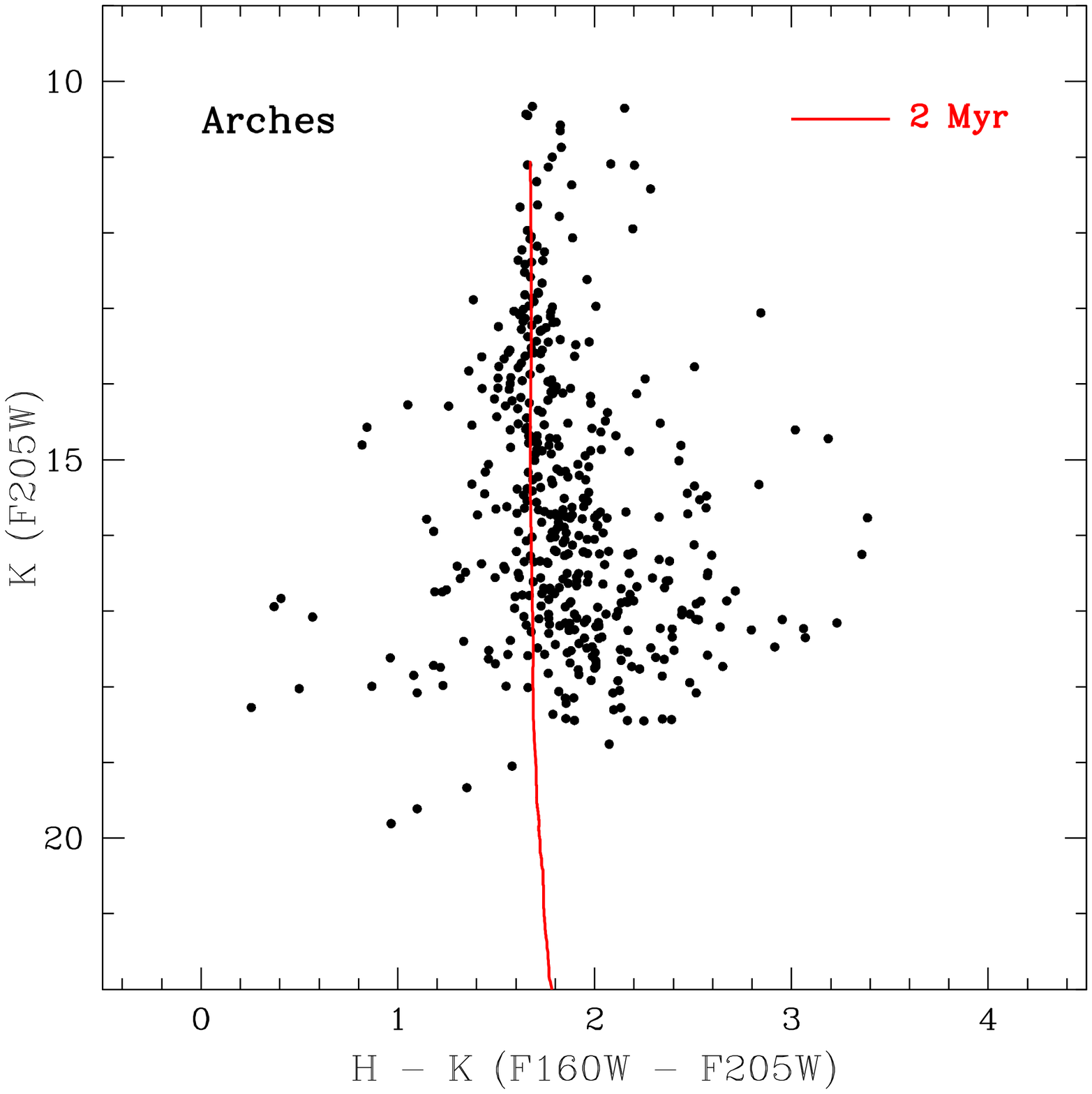}} 
\label{cmd}
\vskip 0.5cm
\begin{center}
\begin{minipage}{0.9\textwidth}
{\bf Fig. 3.}---(Left) Color-magnitude diagram of the Arches cluster
obtained from the Gemini data (filled circles) in comparison with 
$HST/NICMOS$ results given by Figer et al (1999) (open circles).
(Right) Isochrone fits to the color-magnitude diagram of the Arches cluster.
The solid line represents a 2 Myr isochrone from the Geneva models.
\end{minipage}
\end{center}
\end{figure*}
which corresponds to the upper main sequence of massive stars,
indicating a very young age for the Arches cluster.
The mean color of the main sequence is $(H-K)\approx 1.8$.
Almost all the stars in the Gemini field are thought to be  on the main
 sequence.
The main sequence appears to be broadened due to a large amount of differential
reddening  that depends on the variable local extinction.
For comparison, we have also plotted in Fig. 3 the $HST/NICMOS$ photometry
given by Figer et al (1999) for the same field (marked by the square in Fig. 1.).
Our CMD shows in general a good agreement with the $HST/NICMOS$ results.
However, the Gemini photometry is about 2 magnitude shallower than the HST
 photometry
due to the shorter exposures of the Gemini $H$ images.

In this study we adopt a distance modulus of
$(m-M)_0 = -14.52$ (=8 kpc; Reid 1993) for the Arches cluster.
For extinction correction, we use the mean color of the O-type stars in the
 cluster.
We estimate the average color of the observed O-type star candidates with
$12.0< K <15.0$ on the main sequence, obtaining
a value of $(H-K)=1.662$. Comparing this with the
intrinsic  color for O stars, $(H-K)_0 =-0.05$ (Panagia 1973),
we  derive an average color excess
$E(H-K) = (H-K)-(H-K)_0  = 1.662-(-0.05) = 1.712$.
Using the extinction law of Rieke, Rieke, \& Paul (1989), we derive an
extinction value, $A_K = 1.95 E(H-K) = 3.34$.

Then the age of the Arches cluster is estimated using the isochrones given by
the Geneva group.
Using the isochrones with $Z=2 Z_\odot$
and enhanced mass-loss rate
(Meynet et al 1994; Lejeune \& Schaerer 2001),
we derive approximately an age of
$t_{age} \approx 2\pm1~Myr$, as shown in the right panel of Fig. 3.
This value is basically the same as  the value given by Figer et al (1999).

\section{LUMINOSITY FUNCTION}

We have derived the luminosity function of the measured stars in the Arches
 cluster.
To estimate incompleteness of our photometry,
we have carried out the completeness test as follows.
After adding artificial stars into the original frames,
we have analyzed the resulting artificial frames in the same fashion as we applied
to derive the photometry of the Arches cluster in the original images.
We added only 50 stars to each image not to degrade the quality
of the original images, and repeated the same process  to create 400 artificial
frames for each of $H$ and $K'$.
Our experiment was designed that the artificial stars follow random spatial distributions
in each frame and that the LFs of the artificial stars follow the power law distributions.
Table 2 lists the completeness of our photometry we derived thus.
Table 2 shows that our photometry is more than 50\% complete for $K<18$ and $H<19$.
      \begin{table} [!htb]
      \begin{center}
      {\bf Table 2.}  Completeness of our photometry\\
      \begin{tabular}{ c c c }
      \hline \hline
        magnitude   &     F160W &     F205W \\
        \hline
        12.5  &     1.00  &     1.00  \\
        13    &     1.00  &     0.99  \\
        13.5  &     1.00  &     0.99  \\
        14    &     0.98  &     0.94  \\
        14.5  &     0.94  &     0.93  \\
        15    &     0.97  &     0.92  \\
        15.5  &     0.92  &     0.88  \\
        16    &     0.89  &     0.82  \\
        16.5  &     0.86  &     0.75  \\
        17    &     0.86  &     0.71  \\
        17.5  &     0.77  &     0.60  \\
        18    &     0.74  &     0.51  \\
        18.5  &     0.65  &     0.41  \\
        19    &     0.53  &     0.29  \\
        19.5  &     0.39  &     0.18  \\
        20    &     0.27  &     0.13  \\
        20.5  &     0.16  &     0.05  \\
        21    &     0.09  &     0.01  \\
      \hline
      \end{tabular}
      \end{center}
      \label{completeness}
      \end{table}

Since $K'$ images are much deeper than $H$ images, we derive the luminosity functions
of stars from each of $K'$ and $H$ photometry, rather than from the combination
 of both.
We assume the mean color of the stars detected only in the $K'$ images to be
$(H-K)=1.86$.
Fig. 4 displays the $K$ and $H$ LFs derived
from the Gemini images before incompleteness correction.
The LFs derived from the $NICMOS$ images for the same
\begin{figure*}[!htb]
\centerline{\epsfxsize=0.9\textwidth\epsfbox{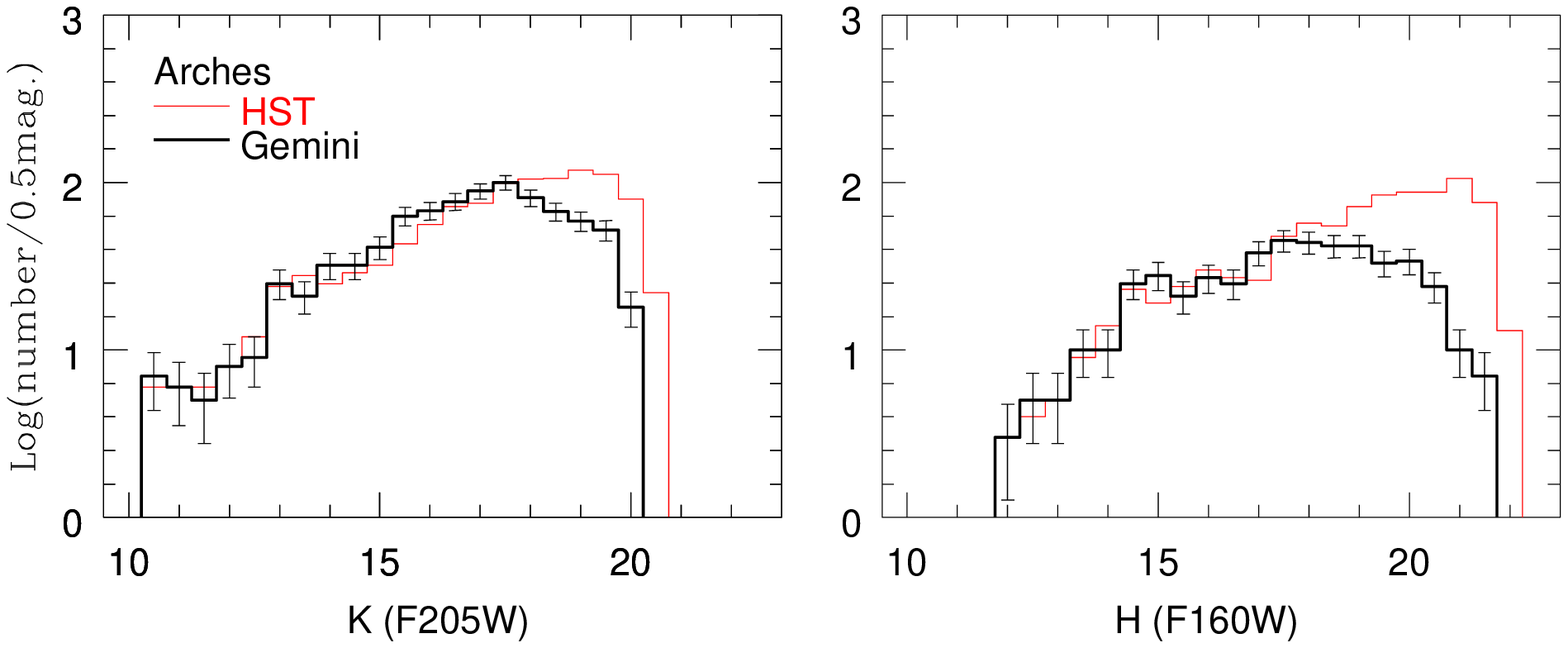}} 
\begin{center}
{\begin{minipage}{0.9\textwidth}
{\bf Fig. 4.}---~~Luminosity functions of the Arches cluster derived
from the Gemini $H$ and $K'$ images before incompleteness correction
(thick lines with error bars).
Light lines represent the LFs derived from the $NICMOS$ data for the
same area of the cluster.
\end{minipage}}
\end{center}
\end{figure*}
 region as the Gemini field
are also plotted for comparison.
Fig. 4 shows that the LFs increase up to  $K\approx 18$ and $H \approx 18$,
and start decreasing thereafter. 
The turnover
appears to be due to incompleteness of the Gemini photometry.
It is found in Fig. 4 that the Gemini LFs are very similar to the NICMOS LFs for
 $K<18$ and $H<18$.

Since the Gemini field is so small that there is little area which can be used
as a control field for deriving the LFs of the cluster.
Therefore we have used the $NICMOS$ data for deriving the LFs of the control field
which can be used for estimation of the field stars in the Gemini data.
Although the characteristics of the Gemini data and the $NICMOS$ data are
 different,
the incompleteness-corrected LFs can be used approximately as a useful guide.
We selected  the outer regions at $r>9''.5$ from the center of
the cluster in the $NICMOS$ images as a control field.
The ratio of the area of the cluster field to that of the control field is 2.77,
which is used for normalizing the LFs of the cluster region.
The data for the incompleteness of the $NICMOS$ photometry were provided by Figer et al(1999).
Fig. 5 (Upper panels) shows the incompleteness-corrected LFs of the control
 field thus derived and the LFs of the Arches cluster (thick lines with error bars).
In the lower panels of Fig. 5 we display the LFs of the Arches cluster after
incompleteness correction and field star subtraction.
We plot 
the LFs for $K<18$ and $H<19$ for which the completeness in our
photometry is higher than 50 \% in Fig. 5.

\section{INITIAL MASS FUNCTION}

Luminosity functions of stars in clusters can be used
to derive the initial mass functions with which stars in clusters form.
For deriving the IMF we have followed the similar procedures
to those used by Figer et al (1999).
We have converted the $K$-band LF derived in the previous section to the IMF,
using the Geneva isochrone with an age of 2 Myrs as derived above.
Fig. 6 shows the resulting IMF of the Arches cluster.
Fitting a single power-law to the data, we derive a
value for the slope of the IMF,
 $\Gamma = -0.79 \pm0.16$ for the range of $1.0< \log~ (M/M_\odot) <2.1$.
This value is in good agreement with the result
based on the $HST/NICMOS$ data, $\Gamma_{0.8-2.1} = -0.7 \pm0.1$,
derived by Figer et al (1999).
This result confirms that the
\begin{figure*} [!htb]
\vskip -0.5cm
\centerline{\epsfxsize=0.9\textwidth\epsfbox{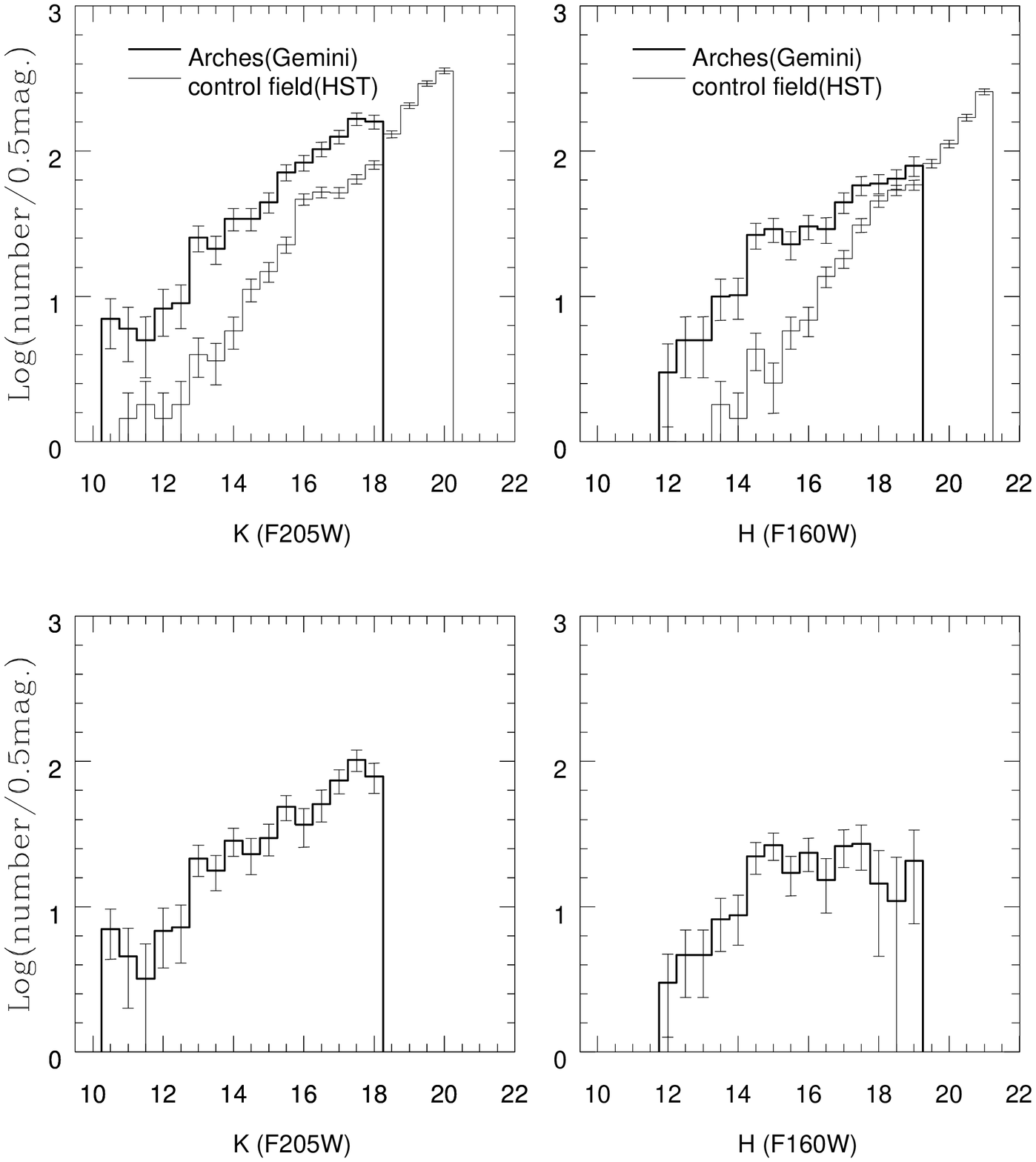}}
\vskip -0.5cm
\vskip -1.5cm
\begin{center}
{\begin{minipage}{0.9\textwidth}
{\bf Fig. 5.}---~~Luminosity functions of the Arches cluster
after incompleteness correction.
(Upper panels)  The LFs of the Arches cluster derived from the Gemini $H$ and $K'$ images
(thick lines with error bars). The LFs of the control field derived from the $NICMOS$ data
with the same area as the Gemini field are also plotted for comparison.
(Lower panels) The LFs of the Arches cluster after subtraction of field contamination
using the $NICMOS$ data for the control field.
\end{minipage}}
\end{center}
\end{figure*}
IMF of the Arches cluster is indeed  much flatter than that of the solar neighborhood which has
an average value between $\Gamma =-1$ and $\Gamma =-2$ (Scalo 1998). 

This result is also consistent with a recent
 finding based on
the $HST/WFPC2$ data
that the IMFs of the bright stars in young clusters in M33
get flatter as the galactocentric distance decreases (Lee et al. 2001).
This trend for the IMF depending on the galactocentric radius is not
completely understood by any single theory until now, requiring detailed
theoretical studies in the future.
However, it may be explained
by the photoevaporative process which
provides a viable mechanism for ablating
massive protostellar cores. In a dense environment
where mass segregation occurs, massive stars in the more metal-rich
center suffer less from ablation than low-mass stars. As a result,
the IMFs get steeper, as the galactocentric radius increases
and as the metallicity decreases (see  
Waller et al 2002 for details).
This result leads to a prediction that the most top-heavy (the flattest)
IMFs may occur near the metal-rich centers of star-forming galaxies
(Lee et al 2001).

\begin{figure} [!hbt]
\epsfxsize=0.45\textwidth\epsfbox{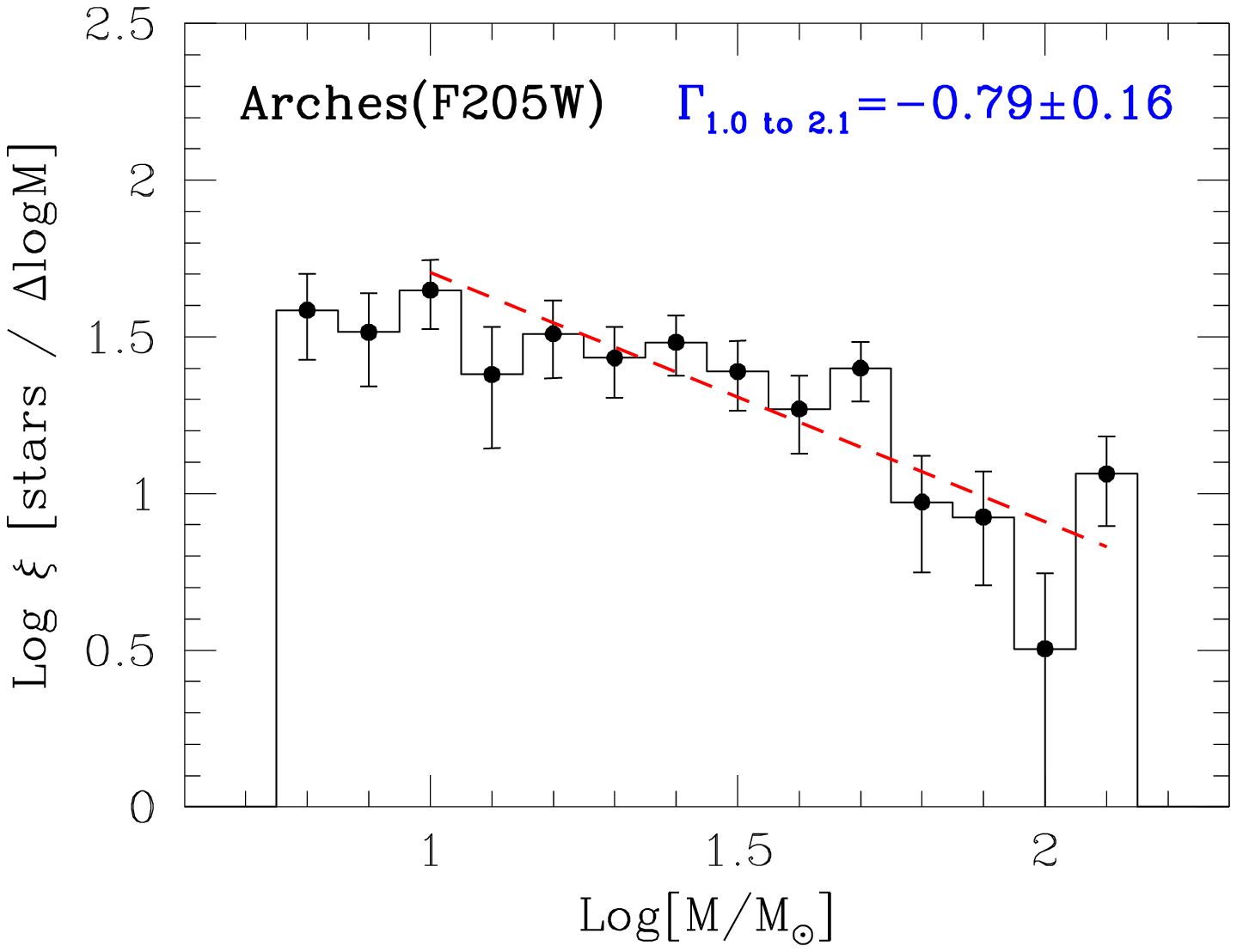} 
\label{IMF}
\vskip 0.5cm
\hskip 0.1\textwidth {\begin{minipage}{0.45\textwidth}
{\bf Fig. 6.}---~~Initial mass function of the Arches cluster derived from
the $K$ images. 
The dashed line represents a fit to the data for the range of
$1.0<\log~(M/M_\odot) <2.1$.
\end{minipage}}
\end{figure}

\section{SUMMARY}

We present Near-IR photometry of the Arches cluster near the Galactic center,
using the data obtained for scientific demonstration with the Gemini/AO.
Primary results are summarized as follows:
First, the color-magnitude diagram of the Arches cluster shows a dominant
blue main sequence consisting mainly of massive stars.
Second, the age of the Arches cluster is estimated to be $2\pm1$ Myrs,
using the Geneva isochrones. This value is consistent with that based
on the $HST/NICMOS$ data (Figer et al 1999).
Third, the $K$ and $H$ luminosity functions of the bright stars in the Arches
 clusters are derived,
showing a slow increase toward the faint end.
Fourth, the initial mass function of the massive stars with
$1.0< \log~ (M/M_\odot) <2.1$ is derived.
Fitting the power law to the data, we obtain a value for
the IMF slope, $\Gamma = -0.79\pm 0.16$.
This confirms that Figer et al (1999)'s result
that the IMF of the Arches cluster is much flatter than that of the solar
neighborhood.
Although the $HST$ provides unprecedented spatial resolution for the dense region,
our results
show that the ground-based AO systems in the Gemini telescope can yield
comparable scientific results with some limits.

\acknowledgments{
Based on observations obtained at the Gemini Observatory, which is
operated by the Association of Universities for Research in Astronomy,
Inc., under a cooperative agreement with the NSF on behalf of the
Gemini partnership: the National Science Foundation (United States),
the Particle Physics and Astronomy Research Council (United Kingdom),
the National Research Council (Canada), CONICYT (Chile), the
Australian Research Council (Australia), CNPq (Brazil) and CONICET
(Argentina). We wish to thank Donald F. Figer for providing the $NICMOS$
 photometry data and
helpful comments.
The authors are grateful to the anonymous referee for useful suggestions.
This work was supported in part by the Korea Research Foundation Grant
 (KRF-2000-DP0450) (to MGL).
}

\input{tableA1}
%
\end{document}

%% file: tableA1.tex
\small
\begin{table}[!htp]
\begin{center}
{\bf Tablel A1.} $HK$ Photometry of the Arches cluster \\
\tabcolsep=1mm
\begin{tabular}[b]{ rrrrrrr }
\hline \hline
ID	&	X  	&	Y  	&$m_{\mathrm{{\scriptscriptstyle F160W}}}$&$\sigma_{\mathrm{\scriptscriptstyle F160W}}$
&$m_{\mathrm{\scriptscriptstyle F205W}}$&$\sigma_{\mathrm{\scriptscriptstyle F205W}}$	\\
\hline
1	&	13.31 	&	798.29 	&	18.458 	&	0.018 	&	16.549 	&	0.007 	\\
2	&	13.48 	&	877.90 	&	18.203 	&	0.015 	&	16.239 	&	0.007 	\\
3	&	24.36 	&	299.74 	&	19.408 	&	0.032 	&	16.868 	&	0.016 	\\
4	&	25.32 	&	106.45 	&	19.422 	&	0.030 	&	16.906 	&	0.010 	\\
5	&	29.01 	&	550.91 	&	19.154 	&	0.019 	&	15.768 	&	0.004 	\\
6	&	65.87 	&	466.82 	&	14.142 	&	0.001 	&	11.948 	&	0.001 	\\
7	&	66.83 	&	186.00 	&	19.449 	&	0.026 	&	16.734 	&	0.007 	\\
8	&	85.05 	&	964.18 	&	13.705 	&	0.005 	&	11.420 	&	0.001 	\\
9	&	85.82 	&	613.33 	&	17.312 	&	0.009 	&	16.941 	&	0.006 	\\
10	&	90.75 	&	915.24 	&	18.510 	&	0.020 	&	16.689 	&	0.010 	\\
11	&	113.81 	&	878.84 	&	18.015 	&	0.009 	&	16.128 	&	0.004 	\\
12	&	116.73 	&	604.14 	&	18.966 	&	0.018 	&	16.601 	&	0.005 	\\
13	&	135.46 	&	900.71 	&	18.685 	&	0.019 	&	16.642 	&	0.009 	\\
14	&	137.13 	&	159.57 	&	18.959 	&	0.019 	&	16.778 	&	0.006 	\\
15	&	141.28 	&	628.29 	&	18.011 	&	0.008 	&	15.967 	&	0.003 	\\
16	&	155.80 	&	884.66 	&	18.416 	&	0.012 	&	16.577 	&	0.006 	\\
17	&	160.11 	&	577.34 	&	18.063 	&	0.008 	&	15.529 	&	0.003 	\\
18	&	167.72 	&	667.00 	&	18.362 	&	0.011 	&	16.502 	&	0.004 	\\
19	&	172.04 	&	867.77 	&	19.025 	&	0.019 	&	16.890 	&	0.006 	\\
20	&	174.60 	&	85.83 	&	18.894 	&	0.017 	&	16.678 	&	0.005 	\\
21	&	180.03 	&	745.25 	&	17.703 	&	0.007 	&	16.021 	&	0.004 	\\
22	&	195.30 	&	495.61 	&	18.630 	&	0.011 	&	16.125 	&	0.003 	\\
23	&	215.71 	&	370.56 	&	17.908 	&	0.006 	&	14.721 	&	0.002 	\\
24	&	217.64 	&	141.69 	&	19.063 	&	0.021 	&	16.864 	&	0.006 	\\
25	&	227.92 	&	639.15 	&	18.432 	&	0.010 	&	16.258 	&	0.004 	\\
26	&	241.55 	&	677.38 	&	18.646 	&	0.015 	&	16.318 	&	0.005 	\\
27	&	246.42 	&	388.05 	&	18.769 	&	0.020 	&	16.894 	&	0.011 	\\
28	&	248.56 	&	784.94 	&	15.531 	&	0.001 	&	13.633 	&	0.001 	\\
29	&	248.79 	&	662.92 	&	17.004 	&	0.005 	&	15.152 	&	0.003 	\\
30	&	251.56 	&	211.32 	&	17.739 	&	0.006 	&	15.725 	&	0.003 	\\
31	&	263.63 	&	365.80 	&	18.269 	&	0.014 	&	16.244 	&	0.005 	\\
32	&	266.06 	&	517.45 	&	16.571 	&	0.003 	&	14.585 	&	0.002 	\\
33	&	282.63 	&	997.29 	&	17.880 	&	0.009 	&	15.864 	&	0.004 	\\
34	&	282.96 	&	685.03 	&	18.462 	&	0.012 	&	16.794 	&	0.006 	\\
35	&	285.46 	&	246.33 	&	18.416 	&	0.011 	&	16.247 	&	0.004 	\\
36	&	285.47 	&	177.83 	&	17.833 	&	0.008 	&	15.770 	&	0.003 	\\
37	&	297.76 	&	404.94 	&	18.282 	&	0.013 	&	16.211 	&	0.007 	\\
38	&	304.65 	&	964.61 	&	18.300 	&	0.012 	&	16.613 	&	0.006 	\\
39	&	310.57 	&	333.01 	&	13.173 	&	0.003 	&	11.091 	&	0.002 	\\
40	&	313.51 	&	818.23 	&	17.643 	&	0.006 	&	15.827 	&	0.003 	\\
41	&	318.94 	&	658.65 	&	16.705 	&	0.005 	&	15.009 	&	0.003 	\\
42	&	325.35 	&	678.74 	&	17.557 	&	0.008 	&	15.826 	&	0.005 	\\
43	&	346.69 	&	639.08 	&	16.350 	&	0.003 	&	14.685 	&	0.002 	\\
44	&	351.02 	&	410.35 	&	18.756 	&	0.018 	&	16.877 	&	0.008 	\\
45	&	356.20 	&	985.03 	&	17.220 	&	0.006 	&	15.265 	&	0.003 	\\
46	&	357.48 	&	323.85 	&	14.579 	&	0.003 	&	12.618 	&	0.002 	\\
47	&	359.89 	&	313.05 	&	17.509 	&	0.016 	&	15.545 	&	0.008 	\\
48	&	363.00 	&	592.65 	&	16.355 	&	0.004 	&	14.683 	&	0.002 	\\
49	&	363.23 	&	583.24 	&	17.094 	&	0.008 	&	15.411 	&	0.003 	\\
50	&	364.02 	&	77.42 	&	18.556 	&	0.025 	&	16.962 	&	0.018 	\\
51	&	365.38 	&	938.78 	&	17.944 	&	0.008 	&	16.270 	&	0.004 	\\
52	&	373.63 	&	895.37 	&	17.725 	&	0.008 	&	16.072 	&	0.004 	\\
53	&	385.82 	&	902.26 	&	16.100 	&	0.004 	&	14.446 	&	0.003 	\\
54	&	388.88 	&	750.41 	&	17.500 	&	0.005 	&	15.661 	&	0.002 	\\
55	&	390.60 	&	97.36 	&	15.281 	&	0.002 	&	13.768 	&	0.001 	\\
56	&	393.77 	&	567.25 	&	17.133 	&	0.018 	&	15.728 	&	0.008 	\\
57	&	397.01 	&	347.02 	&	18.576 	&	0.026 	&	16.667 	&	0.016 	\\
58	&	400.25 	&	66.41 	&	18.856 	&	0.144 	&	16.562 	&	0.106 	\\
59	&	401.67 	&	485.62 	&	15.066 	&	0.128 	&	13.640 	&	0.094 	\\
60	&	402.46 	&	690.07 	&	17.498 	&	0.006 	&	15.722 	&	0.003 	\\
\hline
\end{tabular}
\end{center}
\end{table}

\begin{table}[!htb]
\begin{center}
{\bf Tablel A1.} continued.\\
\tabcolsep=1mm
\begin{tabular}{ rrrrrrr }
\hline \hline
ID    &     X     &     Y     &$m_{\mathrm{{\scriptscriptstyle F160W}}}$&$\sigma_{\mathrm{\scriptscriptstyle F160W}}$
&$m_{\mathrm{\scriptscriptstyle F205W}}$&$\sigma_{\mathrm{\scriptscriptstyle F205W}}$     \\
\hline
61	&	410.87 	&	992.19 	&	17.896 	&	0.010 	&	15.882 	&	0.005 	\\
62	&	422.47 	&	940.35 	&	19.051 	&	0.022 	&	16.884 	&	0.007 	\\
63	&	423.33 	&	871.85 	&	18.112 	&	0.012 	&	16.500 	&	0.005 	\\
64	&	428.48 	&	865.46 	&	18.795 	&	0.022 	&	16.944 	&	0.008 	\\
65	&	429.51 	&	233.92 	&	17.552 	&	0.022 	&	15.743 	&	0.012 	\\
66	&	431.58 	&	128.76 	&	18.436 	&	0.015 	&	16.385 	&	0.006 	\\
67	&	436.16 	&	678.00 	&	18.490 	&	0.019 	&	16.523 	&	0.010 	\\
68	&	439.35 	&	862.51 	&	18.422 	&	0.015 	&	16.503 	&	0.008 	\\
69	&	442.54 	&	615.97 	&	17.357 	&	0.006 	&	15.511 	&	0.003 	\\
70	&	446.08 	&	278.43 	&	13.310 	&	0.003 	&	11.108 	&	0.002 	\\
71	&	447.95 	&	766.26 	&	18.507 	&	0.019 	&	16.729 	&	0.011 	\\
72	&	449.92 	&	990.33 	&	17.442 	&	0.006 	&	15.013 	&	0.003 	\\
73	&	455.22 	&	107.64 	&	16.278 	&	0.003 	&	13.771 	&	0.001 	\\
74	&	455.44 	&	667.17 	&	14.880 	&	0.003 	&	13.104 	&	0.002 	\\
75	&	459.78 	&	256.61 	&	16.637 	&	0.049 	&	14.818 	&	0.035 	\\
76	&	461.76 	&	770.70 	&	15.517 	&	0.004 	&	13.794 	&	0.003 	\\
77	&	462.74 	&	281.99 	&	16.341 	&	0.018 	&	14.128 	&	0.012 	\\
78	&	462.89 	&	703.99 	&	18.661 	&	0.020 	&	16.932 	&	0.012 	\\
79	&	463.34 	&	153.75 	&	17.249 	&	0.006 	&	14.810 	&	0.003 	\\
80	&	475.28 	&	754.90 	&	18.405 	&	0.027 	&	16.809 	&	0.016 	\\
81	&	475.72 	&	235.36 	&	16.899 	&	0.020 	&	14.947 	&	0.014 	\\
82	&	479.80 	&	657.60 	&	16.931 	&	0.006 	&	15.122 	&	0.004 	\\
83	&	480.31 	&	175.97 	&	16.900 	&	0.005 	&	14.866 	&	0.003 	\\
84	&	482.30 	&	294.59 	&	14.978 	&	0.003 	&	12.971 	&	0.002 	\\
85	&	484.29 	&	773.87 	&	15.774 	&	0.003 	&	13.992 	&	0.002 	\\
86	&	492.61 	&	333.23 	&	17.854 	&	0.009 	&	15.347 	&	0.005 	\\
87	&	493.89 	&	989.12 	&	18.565 	&	0.015 	&	16.769 	&	0.007 	\\
88	&	495.94 	&	677.95 	&	17.807 	&	0.009 	&	16.031 	&	0.005 	\\
89	&	504.33 	&	939.50 	&	19.034 	&	0.020 	&	16.876 	&	0.007 	\\
90	&	508.58 	&	126.87 	&	17.626 	&	0.005 	&	14.606 	&	0.002 	\\
91	&	511.99 	&	825.72 	&	15.284 	&	0.001 	&	13.550 	&	0.001 	\\
92	&	514.87 	&	151.05 	&	18.162 	&	0.009 	&	15.327 	&	0.003 	\\
93	&	516.07 	&	613.71 	&	15.279 	&	0.003 	&	13.585 	&	0.002 	\\
94	&	521.21 	&	741.74 	&	15.416 	&	0.003 	&	13.443 	&	0.002 	\\
95	&	524.83 	&	255.82 	&	16.234 	&	0.004 	&	14.253 	&	0.003 	\\
96	&	530.13 	&	570.78 	&	14.903 	&	0.003 	&	13.223 	&	0.002 	\\
97	&	533.13 	&	465.32 	&	17.722 	&	0.018 	&	15.693 	&	0.012 	\\
98	&	533.49 	&	738.87 	&	18.047 	&	0.009 	&	15.478 	&	0.003 	\\
99	&	536.74 	&	488.22 	&	17.266 	&	0.006 	&	15.562 	&	0.003 	\\
100	&	543.28 	&	440.70 	&	13.604 	&	0.001 	&	11.783 	&	0.001 	\\
101	&	544.92 	&	209.91 	&	17.922 	&	0.008 	&	16.002 	&	0.004 	\\
102	&	551.63 	&	764.89 	&	18.464 	&	0.014 	&	16.718 	&	0.006 	\\
103	&	553.40 	&	26.72 	&	16.851 	&	0.006 	&	14.517 	&	0.003 	\\
104	&	553.84 	&	297.10 	&	13.250 	&	0.003 	&	11.366 	&	0.002 	\\
105	&	555.33 	&	677.58 	&	17.743 	&	0.007 	&	15.943 	&	0.003 	\\
106	&	558.03 	&	248.77 	&	18.130 	&	0.035 	&	16.367 	&	0.015 	\\
107	&	562.75 	&	339.36 	&	15.237 	&	0.007 	&	13.412 	&	0.005 	\\
108	&	569.55 	&	979.09 	&	17.198 	&	0.006 	&	15.542 	&	0.003 	\\
109	&	575.94 	&	586.78 	&	14.399 	&	0.003 	&	12.666 	&	0.002 	\\
110	&	580.95 	&	275.17 	&	15.934 	&	0.021 	&	14.056 	&	0.015 	\\
111	&	581.11 	&	484.14 	&	18.286 	&	0.016 	&	16.561 	&	0.009 	\\
112	&	584.44 	&	500.77 	&	18.471 	&	0.018 	&	16.698 	&	0.010 	\\
113	&	593.32 	&	347.65 	&	18.050 	&	0.017 	&	16.051 	&	0.009 	\\
114	&	594.23 	&	780.81 	&	18.438 	&	0.013 	&	16.697 	&	0.006 	\\
115	&	601.06 	&	265.02 	&	13.955 	&	0.003 	&	12.067 	&	0.002 	\\
116	&	601.92 	&	183.67 	&	17.617 	&	0.006 	&	15.731 	&	0.003 	\\
117	&	605.02 	&	351.85 	&	17.239 	&	0.019 	&	16.833 	&	0.013 	\\
118	&	606.15 	&	635.89 	&	16.468 	&	0.004 	&	14.767 	&	0.002 	\\
119	&	606.85 	&	666.02 	&	15.802 	&	0.007 	&	14.177 	&	0.004 	\\
120	&	616.94 	&	217.67 	&	17.823 	&	0.011 	&	15.967 	&	0.005 	\\
\hline
\end{tabular}
\end{center}
\end{table}

\begin{table}[!htb]
\begin{center}
{\bf Tablel A1.} continued.\\
\tabcolsep=1mm
\begin{tabular}{ rrrrrrr }
\hline \hline
ID    &     X     &     Y     &$m_{\mathrm{{\scriptscriptstyle F160W}}}$&$\sigma_{\mathrm{\scriptscriptstyle F160W}}$
&$m_{\mathrm{\scriptscriptstyle F205W}}$&$\sigma_{\mathrm{\scriptscriptstyle F205W}}$     \\
\hline
121	&	618.52 	&	438.86 	&	16.282 	&	0.003 	&	14.538 	&	0.002 	\\
122	&	624.19 	&	196.86 	&	19.538 	&	0.027 	&	16.866 	&	0.007 	\\
123	&	624.90 	&	771.86 	&	16.392 	&	0.016 	&	14.684 	&	0.009 	\\
124	&	626.30 	&	536.89 	&	16.543 	&	0.081 	&	14.488 	&	0.058 	\\
125	&	629.25 	&	390.83 	&	18.839 	&	0.027 	&	16.704 	&	0.016 	\\
126	&	631.96 	&	903.09 	&	17.915 	&	0.008 	&	15.444 	&	0.004 	\\
127	&	633.77 	&	462.13 	&	17.800 	&	0.028 	&	16.375 	&	0.015 	\\
128	&	637.38 	&	747.37 	&	18.173 	&	0.027 	&	16.556 	&	0.013 	\\
129	&	637.52 	&	408.21 	&	18.502 	&	0.016 	&	16.770 	&	0.009 	\\
130	&	638.40 	&	447.58 	&	17.514 	&	0.010 	&	15.629 	&	0.006 	\\
131	&	639.51 	&	478.27 	&	14.976 	&	0.013 	&	13.188 	&	0.008 	\\
132	&	640.88 	&	684.52 	&	17.401 	&	0.019 	&	15.431 	&	0.007 	\\
133	&	644.45 	&	99.42 	&	18.188 	&	0.009 	&	15.715 	&	0.003 	\\
134	&	646.61 	&	510.09 	&	12.402 	&	0.016 	&	10.576 	&	0.012 	\\
135	&	646.89 	&	588.43 	&	17.991 	&	0.014 	&	16.196 	&	0.008 	\\
136	&	648.01 	&	569.27 	&	17.377 	&	0.008 	&	15.663 	&	0.005 	\\
137	&	650.17 	&	857.43 	&	14.825 	&	0.004 	&	13.048 	&	0.003 	\\
138	&	653.28 	&	693.10 	&	14.166 	&	0.003 	&	12.522 	&	0.002 	\\
139	&	653.55 	&	742.92 	&	17.947 	&	0.031 	&	16.408 	&	0.013 	\\
140	&	654.98 	&	787.46 	&	13.729 	&	0.004 	&	12.051 	&	0.003 	\\
141	&	655.00 	&	469.41 	&	17.175 	&	0.036 	&	15.621 	&	0.023 	\\
142	&	656.24 	&	29.89 	&	18.720 	&	0.017 	&	16.340 	&	0.006 	\\
143	&	657.38 	&	647.81 	&	18.857 	&	0.025 	&	16.263 	&	0.010 	\\
144	&	659.93 	&	454.88 	&	16.529 	&	0.012 	&	14.722 	&	0.008 	\\
145	&	660.06 	&	270.11 	&	16.702 	&	0.006 	&	14.923 	&	0.003 	\\
146	&	661.27 	&	614.38 	&	19.432 	&	0.060 	&	16.988 	&	0.021 	\\
147	&	667.98 	&	814.95 	&	16.891 	&	0.029 	&	15.450 	&	0.014 	\\
148	&	668.30 	&	943.29 	&	17.514 	&	0.032 	&	15.713 	&	0.022 	\\
149	&	668.86 	&	447.54 	&	18.108 	&	0.020 	&	16.243 	&	0.013 	\\
150	&	671.97 	&	768.86 	&	15.934 	&	0.031 	&	14.431 	&	0.016 	\\
151	&	672.06 	&	526.71 	&	17.283 	&	0.156 	&	15.640 	&	0.111 	\\
152	&	672.31 	&	752.37 	&	16.176 	&	0.041 	&	14.605 	&	0.021 	\\
153	&	676.64 	&	495.21 	&	16.790 	&	0.085 	&	14.682 	&	0.059 	\\
154	&	681.79 	&	80.01 	&	19.105 	&	0.021 	&	16.532 	&	0.006 	\\
155	&	688.05 	&	794.04 	&	15.687 	&	0.021 	&	14.196 	&	0.012 	\\
156	&	692.85 	&	502.65 	&	17.815 	&	0.024 	&	16.212 	&	0.012 	\\
157	&	693.27 	&	759.10 	&	12.784 	&	0.009 	&	10.999 	&	0.006 	\\
158	&	696.37 	&	282.54 	&	18.161 	&	0.009 	&	16.219 	&	0.004 	\\
159	&	697.19 	&	795.21 	&	15.200 	&	0.013 	&	13.523 	&	0.008 	\\
160	&	703.22 	&	722.71 	&	16.976 	&	0.041 	&	15.061 	&	0.013 	\\
161	&	704.72 	&	829.84 	&	15.325 	&	0.005 	&	14.274 	&	0.003 	\\
162	&	707.79 	&	783.73 	&	15.393 	&	0.081 	&	13.781 	&	0.058 	\\
163	&	708.27 	&	648.67 	&	16.487 	&	0.007 	&	14.777 	&	0.003 	\\
164	&	710.86 	&	629.52 	&	18.164 	&	0.021 	&	16.218 	&	0.009 	\\
165	&	711.18 	&	586.78 	&	14.465 	&	0.004 	&	12.820 	&	0.003 	\\
166	&	712.26 	&	361.57 	&	15.976 	&	0.003 	&	14.213 	&	0.002 	\\
167	&	713.32 	&	288.74 	&	19.047 	&	0.022 	&	16.691 	&	0.006 	\\
168	&	715.42 	&	966.03 	&	17.313 	&	0.010 	&	15.708 	&	0.006 	\\
169	&	718.15 	&	616.15 	&	17.764 	&	0.025 	&	15.764 	&	0.010 	\\
170	&	719.81 	&	634.91 	&	16.576 	&	0.009 	&	14.871 	&	0.005 	\\
171	&	720.45 	&	508.98 	&	15.587 	&	0.004 	&	13.955 	&	0.003 	\\
172	&	720.98 	&	413.27 	&	17.711 	&	0.008 	&	15.899 	&	0.004 	\\
173	&	728.83 	&	309.28 	&	17.563 	&	0.009 	&	15.618 	&	0.005 	\\
174	&	729.75 	&	910.69 	&	17.451 	&	0.008 	&	15.508 	&	0.004 	\\
175	&	737.36 	&	964.41 	&	17.739 	&	0.012 	&	15.954 	&	0.007 	\\
176	&	737.48 	&	841.70 	&	17.830 	&	0.033 	&	16.487 	&	0.017 	\\
177	&	738.99 	&	643.67 	&	18.424 	&	0.038 	&	16.230 	&	0.015 	\\
178	&	740.72 	&	534.85 	&	17.129 	&	0.058 	&	15.947 	&	0.024 	\\
179	&	740.83 	&	567.82 	&	15.904 	&	0.021 	&	13.059 	&	0.006 	\\
180	&	748.48 	&	650.27 	&	17.649 	&	0.024 	&	15.773 	&	0.010 	\\
\hline
\end{tabular}
\end{center}
\end{table}

\begin{table}[!htb]
\begin{center}
{\bf Tablel A1.} continued.\\
\tabcolsep=1mm
\begin{tabular}{ rrrrrrr }
\hline \hline
ID    &     X     &     Y     &$m_{\mathrm{{\scriptscriptstyle F160W}}}$&$\sigma_{\mathrm{\scriptscriptstyle F160W}}$
&$m_{\mathrm{\scriptscriptstyle F205W}}$&$\sigma_{\mathrm{\scriptscriptstyle F205W}}$     \\
\hline
181	&	749.07 	&	692.04 	&	16.228 	&	0.005 	&	14.571 	&	0.003 	\\
182	&	749.17 	&	958.63 	&	18.973 	&	0.024 	&	16.597 	&	0.007 	\\
183	&	753.10 	&	817.83 	&	14.627 	&	0.006 	&	13.037 	&	0.004 	\\
184	&	756.08 	&	277.13 	&	15.385 	&	0.002 	&	13.481 	&	0.001 	\\
185	&	757.43 	&	663.90 	&	16.444 	&	0.013 	&	14.779 	&	0.008 	\\
186	&	758.81 	&	646.54 	&	15.543 	&	0.007 	&	13.871 	&	0.004 	\\
187	&	760.51 	&	230.36 	&	18.510 	&	0.017 	&	16.600 	&	0.006 	\\
188	&	763.07 	&	475.47 	&	16.521 	&	0.041 	&	15.060 	&	0.023 	\\
189	&	766.93 	&	560.10 	&	12.478 	&	0.012 	&	10.653 	&	0.009 	\\
190	&	767.16 	&	43.01 	&	17.064 	&	0.008 	&	14.889 	&	0.004 	\\
191	&	767.94 	&	77.74 	&	17.605 	&	0.009 	&	15.751 	&	0.005 	\\
192	&	768.08 	&	738.11 	&	15.635 	&	0.009 	&	14.069 	&	0.004 	\\
193	&	771.23 	&	666.77 	&	15.835 	&	0.014 	&	14.288 	&	0.008 	\\
194	&	771.84 	&	732.37 	&	15.565 	&	0.011 	&	13.994 	&	0.004 	\\
195	&	772.12 	&	349.48 	&	15.840 	&	0.003 	&	14.035 	&	0.002 	\\
196	&	775.10 	&	604.65 	&	15.727 	&	0.013 	&	13.945 	&	0.005 	\\
197	&	777.59 	&	28.75 	&	16.441 	&	0.084 	&	14.376 	&	0.058 	\\
198	&	778.33 	&	252.17 	&	17.129 	&	0.007 	&	15.207 	&	0.004 	\\
199	&	781.91 	&	608.80 	&	15.007 	&	0.007 	&	13.254 	&	0.004 	\\
200	&	783.68 	&	365.22 	&	19.072 	&	0.021 	&	16.496 	&	0.005 	\\
201	&	784.09 	&	893.65 	&	15.034 	&	0.007 	&	13.374 	&	0.004 	\\
202	&	787.71 	&	639.47 	&	13.973 	&	0.011 	&	12.362 	&	0.004 	\\
203	&	787.97 	&	508.09 	&	12.014 	&	0.024 	&	10.330 	&	0.017 	\\
204	&	788.50 	&	813.02 	&	14.778 	&	0.007 	&	13.131 	&	0.004 	\\
205	&	788.52 	&	916.38 	&	17.560 	&	0.023 	&	15.738 	&	0.012 	\\
206	&	790.63 	&	717.61 	&	15.187 	&	0.013 	&	13.826 	&	0.005 	\\
207	&	792.42 	&	309.39 	&	17.937 	&	0.016 	&	16.097 	&	0.010 	\\
208	&	793.41 	&	650.32 	&	15.549 	&	0.041 	&	14.291 	&	0.021 	\\
209	&	800.05 	&	435.34 	&	14.600 	&	0.004 	&	12.908 	&	0.003 	\\
210	&	800.19 	&	137.19 	&	16.486 	&	0.003 	&	14.714 	&	0.002 	\\
211	&	801.25 	&	905.26 	&	17.888 	&	0.045 	&	16.571 	&	0.025 	\\
212	&	801.34 	&	858.01 	&	16.244 	&	0.009 	&	14.595 	&	0.005 	\\
213	&	805.14 	&	649.16 	&	14.270 	&	0.031 	&	12.886 	&	0.021 	\\
214	&	805.31 	&	830.31 	&	15.429 	&	0.009 	&	13.920 	&	0.005 	\\
215	&	810.92 	&	27.20 	&	12.701 	&	0.007 	&	10.871 	&	0.004 	\\
216	&	812.34 	&	819.37 	&	15.206 	&	0.007 	&	13.666 	&	0.004 	\\
217	&	812.64 	&	760.87 	&	16.932 	&	0.042 	&	15.786 	&	0.014 	\\
218	&	812.81 	&	950.65 	&	18.041 	&	0.016 	&	16.350 	&	0.008 	\\
219	&	813.73 	&	574.82 	&	15.322 	&	0.023 	&	13.597 	&	0.010 	\\
220	&	815.14 	&	613.09 	&	16.189 	&	0.040 	&	13.932 	&	0.009 	\\
221	&	817.73 	&	489.09 	&	17.091 	&	0.056 	&	15.227 	&	0.027 	\\
222	&	817.82 	&	834.25 	&	15.144 	&	0.008 	&	13.583 	&	0.005 	\\
223	&	818.71 	&	981.20 	&	16.936 	&	0.009 	&	15.223 	&	0.005 	\\
224	&	820.13 	&	883.59 	&	16.606 	&	0.011 	&	15.162 	&	0.006 	\\
225	&	820.78 	&	645.88 	&	12.110 	&	0.021 	&	10.450 	&	0.015 	\\
226	&	821.65 	&	727.77 	&	13.281 	&	0.005 	&	11.660 	&	0.003 	\\
227	&	822.74 	&	626.38 	&	14.682 	&	0.017 	&	13.060 	&	0.008 	\\
228	&	827.86 	&	500.51 	&	18.111 	&	0.163 	&	16.353 	&	0.107 	\\
229	&	830.03 	&	482.47 	&	17.063 	&	0.017 	&	15.093 	&	0.007 	\\
230	&	830.43 	&	690.31 	&	14.905 	&	0.008 	&	13.277 	&	0.003 	\\
231	&	832.54 	&	428.57 	&	17.966 	&	0.032 	&	16.719 	&	0.012 	\\
232	&	833.20 	&	925.02 	&	18.052 	&	0.016 	&	16.557 	&	0.009 	\\
233	&	839.20 	&	695.70 	&	16.137 	&	0.021 	&	14.525 	&	0.008 	\\
234	&	842.80 	&	794.44 	&	15.882 	&	0.010 	&	14.104 	&	0.005 	\\
235	&	843.15 	&	724.21 	&	15.934 	&	0.033 	&	14.325 	&	0.018 	\\
236	&	843.73 	&	85.36 	&	18.965 	&	0.024 	&	16.949 	&	0.009 	\\
237	&	846.15 	&	499.58 	&	17.972 	&	0.063 	&	16.746 	&	0.044 	\\
238	&	847.47 	&	874.16 	&	17.989 	&	0.029 	&	16.347 	&	0.017 	\\
239	&	849.41 	&	931.26 	&	17.106 	&	0.010 	&	15.466 	&	0.006 	\\
240	&	849.42 	&	831.54 	&	15.487 	&	0.014 	&	13.913 	&	0.009 	\\
\hline
\end{tabular}
\end{center}
\end{table}

\begin{table}[!htb]
\begin{center}
{\bf Tablel A1.} continued.\\
\tabcolsep=1mm
\begin{tabular}{ rrrrrrr }
\hline \hline
ID    &     X     &     Y     &$m_{\mathrm{{\scriptscriptstyle F160W}}}$&$\sigma_{\mathrm{\scriptscriptstyle F160W}}$
&$m_{\mathrm{\scriptscriptstyle F205W}}$&$\sigma_{\mathrm{\scriptscriptstyle F205W}}$     \\
\hline
241	&	849.50 	&	261.85 	&	18.114 	&	0.013 	&	16.266 	&	0.007 	\\
242	&	849.58 	&	791.78 	&	16.993 	&	0.022 	&	15.387 	&	0.010 	\\
243	&	851.58 	&	513.50 	&	16.615 	&	0.010 	&	14.882 	&	0.005 	\\
244	&	853.22 	&	547.72 	&	15.350 	&	0.007 	&	13.722 	&	0.004 	\\
245	&	854.93 	&	589.08 	&	13.030 	&	0.006 	&	11.325 	&	0.004 	\\
246	&	856.95 	&	97.14 	&	17.736 	&	0.009 	&	15.890 	&	0.005 	\\
247	&	858.18 	&	748.40 	&	15.623 	&	0.116 	&	14.805 	&	0.081 	\\
248	&	858.70 	&	502.00 	&	16.622 	&	0.012 	&	14.922 	&	0.007 	\\
249	&	860.44 	&	674.43 	&	15.209 	&	0.006 	&	13.444 	&	0.003 	\\
250	&	861.72 	&	538.09 	&	14.633 	&	0.004 	&	12.966 	&	0.003 	\\
251	&	864.41 	&	471.32 	&	16.409 	&	0.014 	&	14.836 	&	0.005 	\\
252	&	868.72 	&	805.45 	&	16.829 	&	0.022 	&	15.166 	&	0.014 	\\
253	&	871.63 	&	273.18 	&	15.915 	&	0.004 	&	14.129 	&	0.002 	\\
254	&	874.66 	&	446.09 	&	13.998 	&	0.004 	&	12.253 	&	0.003 	\\
255	&	878.82 	&	564.91 	&	17.562 	&	0.114 	&	15.948 	&	0.079 	\\
256	&	880.10 	&	323.68 	&	17.429 	&	0.012 	&	15.684 	&	0.006 	\\
257	&	881.72 	&	368.54 	&	17.907 	&	0.012 	&	16.058 	&	0.004 	\\
258	&	883.71 	&	108.47 	&	18.676 	&	0.019 	&	16.501 	&	0.007 	\\
259	&	883.74 	&	300.12 	&	14.986 	&	0.003 	&	13.182 	&	0.002 	\\
260	&	884.08 	&	743.84 	&	12.083 	&	0.019 	&	10.433 	&	0.014 	\\
261	&	893.80 	&	504.37 	&	18.420 	&	0.074 	&	16.788 	&	0.051 	\\
262	&	896.40 	&	652.41 	&	14.751 	&	0.009 	&	13.240 	&	0.005 	\\
263	&	896.96 	&	180.75 	&	18.019 	&	0.012 	&	16.214 	&	0.007 	\\
264	&	906.31 	&	550.05 	&	17.708 	&	0.051 	&	16.407 	&	0.019 	\\
265	&	906.42 	&	246.61 	&	15.883 	&	0.003 	&	14.084 	&	0.002 	\\
266	&	906.71 	&	486.06 	&	18.086 	&	0.038 	&	15.758 	&	0.010 	\\
267	&	908.12 	&	891.75 	&	17.991 	&	0.014 	&	16.446 	&	0.008 	\\
268	&	910.00 	&	646.73 	&	13.628 	&	0.008 	&	11.969 	&	0.005 	\\
269	&	910.60 	&	60.84 	&	17.819 	&	0.009 	&	16.020 	&	0.005 	\\
270	&	917.22 	&	669.23 	&	16.380 	&	0.031 	&	14.516 	&	0.012 	\\
271	&	919.78 	&	368.80 	&	15.019 	&	0.008 	&	13.288 	&	0.005 	\\
272	&	921.10 	&	629.91 	&	14.064 	&	0.009 	&	12.418 	&	0.005 	\\
273	&	921.43 	&	302.65 	&	17.091 	&	0.007 	&	15.366 	&	0.004 	\\
274	&	923.20 	&	529.83 	&	15.801 	&	0.015 	&	14.221 	&	0.007 	\\
275	&	923.99 	&	706.39 	&	14.258 	&	0.008 	&	12.584 	&	0.005 	\\
276	&	925.19 	&	861.50 	&	16.949 	&	0.010 	&	15.264 	&	0.006 	\\
277	&	926.63 	&	401.45 	&	16.060 	&	0.010 	&	14.347 	&	0.005 	\\
278	&	929.85 	&	149.43 	&	15.845 	&	0.004 	&	14.053 	&	0.003 	\\
279	&	931.02 	&	930.44 	&	18.062 	&	0.015 	&	16.340 	&	0.009 	\\
280	&	932.75 	&	537.29 	&	13.753 	&	0.005 	&	12.081 	&	0.003 	\\
281	&	934.90 	&	368.20 	&	14.514 	&	0.005 	&	12.800 	&	0.003 	\\
282	&	935.24 	&	437.87 	&	17.103 	&	0.022 	&	15.317 	&	0.009 	\\
283	&	936.27 	&	599.77 	&	16.664 	&	0.034 	&	14.633 	&	0.010 	\\
284	&	937.49 	&	453.60 	&	17.037 	&	0.015 	&	15.380 	&	0.009 	\\
285	&	940.69 	&	231.90 	&	18.014 	&	0.012 	&	16.052 	&	0.005 	\\
286	&	942.63 	&	435.84 	&	15.262 	&	0.006 	&	13.581 	&	0.003 	\\
287	&	942.76 	&	397.20 	&	14.546 	&	0.005 	&	12.862 	&	0.003 	\\
288	&	943.69 	&	786.33 	&	15.918 	&	0.018 	&	14.250 	&	0.008 	\\
289	&	944.01 	&	905.80 	&	17.933 	&	0.024 	&	16.745 	&	0.011 	\\
290	&	948.37 	&	627.94 	&	14.066 	&	0.008 	&	12.387 	&	0.005 	\\
291	&	949.03 	&	226.44 	&	16.985 	&	0.007 	&	15.157 	&	0.004 	\\
292	&	949.47 	&	661.04 	&	15.120 	&	0.099 	&	13.551 	&	0.072 	\\
293	&	950.85 	&	566.41 	&	17.719 	&	0.042 	&	15.781 	&	0.016 	\\
294	&	955.49 	&	761.76 	&	15.918 	&	0.026 	&	14.542 	&	0.015 	\\
295	&	956.86 	&	493.98 	&	12.897 	&	0.006 	&	11.132 	&	0.004 	\\
296	&	957.71 	&	691.24 	&	15.138 	&	0.009 	&	13.434 	&	0.005 	\\
297	&	964.21 	&	237.82 	&	16.577 	&	0.005 	&	14.809 	&	0.003 	\\
298	&	969.32 	&	383.76 	&	14.501 	&	0.005 	&	12.788 	&	0.003 	\\
299	&	969.91 	&	773.09 	&	13.857 	&	0.010 	&	12.226 	&	0.006 	\\
300	&	970.23 	&	623.74 	&	14.104 	&	0.008 	&	12.367 	&	0.005 	\\
\hline
\end{tabular}
\end{center}
\end{table}

\begin{table}[!htb]
\begin{center}
{\bf Tablel A1.} continued.\\
\tabcolsep=1mm
\begin{tabular}{ rrrrrrr }
\hline \hline                                                                             
ID    &     X     &     Y     &$m_{\mathrm{{\scriptscriptstyle F160W}}}$&$\sigma_{\mathrm{\scriptscriptstyle F160W}}$
&$m_{\mathrm{\scriptscriptstyle F205W}}$&$\sigma_{\mathrm{\scriptscriptstyle F205W}}$     \\
\hline
301	&	970.40 	&	675.82 	&	15.957 	&	0.013 	&	14.118 	&	0.006 	\\
302	&	970.58 	&	588.81 	&	15.561 	&	0.014 	&	14.053 	&	0.006 	\\
303	&	978.28 	&	632.56 	&	14.704 	&	0.012 	&	13.085 	&	0.006 	\\
304	&	980.00 	&	656.26 	&	15.025 	&	0.013 	&	13.302 	&	0.007 	\\
305	&	980.40 	&	901.56 	&	15.487 	&	0.037 	&	14.059 	&	0.020 	\\
306	&	982.36 	&	573.15 	&	17.849 	&	0.046 	&	15.690 	&	0.017 	\\
307	&	987.45 	&	132.59 	&	19.609 	&	0.031 	&	16.251 	&	0.009 	\\
308	&	988.40 	&	634.39 	&	14.855 	&	0.013 	&	13.144 	&	0.007 	\\
309	&	989.04 	&	806.84 	&	15.414 	&	0.108 	&	14.571 	&	0.075 	\\
310	&	989.66 	&	12.73 	&	18.499 	&	0.023 	&	16.634 	&	0.013 	\\
311	&	991.39 	&	517.55 	&	14.650 	&	0.007 	&	13.011 	&	0.004 	\\
312	&	998.25 	&	862.62 	&	16.698 	&	0.016 	&	15.322 	&	0.007 	\\
313	&	998.83 	&	741.32 	&	13.886 	&	0.010 	&	12.178 	&	0.006 	\\
314	&	998.86 	&	538.89 	&	14.766 	&	0.007 	&	12.981 	&	0.005 	\\
315	&	999.75 	&	122.03 	&	18.580 	&	0.019 	&	16.616 	&	0.011 	\\
316	&	1002.16 	&	694.80 	&	15.729 	&	0.098 	&	13.966 	&	0.072 	\\
317	&	1006.10 	&	311.77 	&	17.149 	&	0.028 	&	15.651 	&	0.016 	\\
318	&	1006.72 	&	793.98 	&	12.763 	&	0.012 	&	11.104 	&	0.008 	\\
319	&	1012.71 	&	642.70 	&	13.342 	&	0.009 	&	11.632 	&	0.006 	\\
320	&	1013.21 	&	912.36 	&	12.507 	&	0.022 	&	10.354 	&	0.015 	\\
321	&	1013.92 	&	370.78 	&	16.861 	&	0.013 	&	14.881 	&	0.005 	\\
322	&	1019.91 	&	266.30 	&	16.104 	&	0.008 	&	14.372 	&	0.004 	\\
323	&	1021.98 	&	167.92 	&	17.561 	&	0.010 	&	15.726 	&	0.006 	\\
324	&	1025.87 	&	364.09 	&	14.810 	&	0.006 	&	13.173 	&	0.004 	\\
325	&	1027.45 	&	510.86 	&	16.142 	&	0.012 	&	14.163 	&	0.005 	\\
326	&	1028.92 	&	427.06 	&	17.048 	&	0.009 	&	15.267 	&	0.005 	\\
327	&	1033.92 	&	158.94 	&	18.613 	&	0.032 	&	16.846 	&	0.015 	\\
\hline													
\end{tabular}
\end{center}
\end{table}

%% file: yjyang.bbl
\begin{references}

\reference{}Blum, R. D., Schaerer, D., Pasquali, A., Heydari-Malayeri, M.,
 Conti, P. S.,
\& Schmutz, W. 2001, 2$\mu m$ Narrow-band Adaptive Optics Imaging in the Arches
Cluster, AJ, 122, 1875
\reference{}Figer, D. F., Kim, S. S., Morris, M., Serabyn, E., Rich,
R. M., McLean, I. 1999, $Hubble Space Telescope/NICMOS$ Observations of
Massive Stellar Clusters near the Galactic Center, ApJ, 525, 750

\reference{}Kim, S. S., Figer, D. F., Lee, H. M., Morris, M. 2000, 
N-Body Simulations of Compact Young Clusters near the Galactic Center,
ApJ, 545, 301
\reference{}Lee, M. G., Park, H. S., Kim. S. C., Waller, W. H., Parker, J. W., 
\& Malumuth, E., \& Hodge, P. 2001, $HST-WFPC2$ Observations of the Star Clusters
 in the
Giant HII Regions of M33, astro-ph 0109258
\reference{}Lejeune, T., \& Schaerer, D. 2001, 
Database of Geneva Stellar Evolution Tracks and Isochrones for
UBVRI JHKLL'M, $HST-WFPC2$, Geneva, and Washington systems,
A\&A, 366, 538
\reference{}Meynet, G., Maeder, A., Schaller, G.,Scahaerer, D., \&
Charbonnel, C. 1994, Grids of massive stars with high mass loss rates. V. 
From 12 to 120 $M_{\odot}$ at Z=0.001, 0.004, 0.008, 0.020 and 0.040, A\&AS,
 103, 97
\reference{}Nagata, T., Woodward, C. E., Shure, M., \& Kobayashi, N. 1995, 
Object 17: Another Cluster of Emission-Line Stars Near the Galactic Center,
AJ, 109, 1676 
\reference{}Panagia, N. 1973, Some Physical parameters of early-type stars, AJ,
 78, 929 
\reference{}Reid, M. J. 1993, The distance to the center of the Galaxy, ARA\&A,
 31, 345
\reference{}Rieke, G. H.,Rieke, M. J., \& Paul, A. E. 1989, Origin of
the excitation of the galactic center, ApJ, 336, 752
\reference{}Salpeter, E. E. 1955, The Luminosity Function and Stellar Evolution,
 ApJ, 121, 161
\reference{}Scalo, J. 1998, The IMF Revisted: A Case for Variations, ASP Conf.
 Ser., 142, 
The Stellar Initial Mass Function, ed., G. Gilmore \& D. Howell, 201
\reference{}Stetson, P. 1987, DAOPHOT - A computer program for crowded-field
 stellar photometry, PASP, 99, 191
\reference{}Stolte, A., Grebel, E. K., Brandner, W., \& Figer, D. F. 2002, 
 The Mass Function of the Arches Cluster from Gemini Adaptive Optics
Data, astro-ph 0206360
\reference{} Waller, W. H., Lee, M. G., Park, H. S., Kim, S. C. et al 2002,
Systematic IMF Variations in the Giant HII Regions of M33, in preparation
\reference{}Yang, Y., Park, H. S., Lee, M. G., \& Lee, S.-G. 2002, 
Near-IR Photometry of the Arches Cluster close to the Galactic
Center, Bull. of Korean Astro. Soc., 27, 50
\reference{}Yuzef-Zadeh, F., Law, C., Wardle, M., Wang, Q. D., Fruscione, A.,
Lang, C. C., \& Cotera, A. 2002, 
Detection of X-ray Emissioin from the Arches Cluster Near the Galactic Center,
ApJ, 570, 665
\end{references}
